\documentclass{mem}
\usepackage{natbib}
\usepackage{graphicx}
\begin{document}

\title{
Stellar Variability in the Sub-Horizontal Branch Region of the LMC
}

   \subtitle{}

\author{
D. \,Lepischak 
\and D.L.\, Welch
          }

  \offprints{D. Lepischak}

\institute{
Department of Physics and Astronomy,
McMaster University,
Hamilton, Ontario
L8S 4M1, Canada.
\email{lepischak@physics.mcmaster.ca}
}

\authorrunning{Lepischak}

\titlerunning{LMC Sub-HB Variables}

\abstract{ We present the initial results from an in-depth
re-examination of the MACHO project Large Magellanic Cloud database to
identify and characterize stellar variability near the intersection of
the instability strip and the main sequence.  This dataset's long
time-series and uniform photometry is an unprecedented resource for
describing the frequency and regions of incidence of various radial
and non-radial modes of excitation.  The raw MACHO photometry has been
investigated to identify factors responsible for most of the residual
photometric variance and increase the sensitivity to small amplitudes.
We present details of the search method used, the $\delta$ Scuti
variables detected thus far and discuss the implications for the
completed survey.  } 

\maketitle{}

\section{Introduction}
The region of the colour-magnitude diagram where the Cepheid
instability strip approaches and meets the main sequence is populated
by a variety of variable star types.  These objects are difficult to
detect in current LMC surveys: typical amplitudes are less than 0.1
magnitudes, periods are as short as a few hours and they have typical
magnitudes that place them near the photometric detection limit.

Of particular interest are the $\delta$ Scuti pulsators, especially
those of high amplitudes ($A_V \geq 0.1-0.3$) also known as HADS.
Analysis of a significant sample of LMC HADS would provide a valuable
complement to studies of these objects in our Galaxy and nearby dwarf
galaxies and could allow us to:
\begin{itemize}
\item test P-L relations on a sizeable sample at a uniform distance
\item examine the fraction of such stars that show double and multi-mode
pulsation and investigate the period ratios at a different metallicity
\item possibly detect the theoretically predicted period change due to
evolution.  The long time-baseline and uniform coverage of the MACHO
data are ideal for this.
\end{itemize}

\section{MACHO Photometry}

There are a number of factors that contribute additional scatter to
the lightcurves especially at fainter magnitudes.  To compensate we
correlated the residuals and reduction flags for over 15 million
observations defining limits for various flags beyond which the
scatter was too great.  Observations with flags beyond these limits
were excluded.  The most significant factors were found to be
crowding, poor sky subtraction and chunk overlap.  This process
reduced the scatter by a statistically significant amount in 80.8\% of
the systems.

\section{Search and Results}

\begin{figure}[th!]
\resizebox{\columnwidth}{!}{\includegraphics[clip=true]{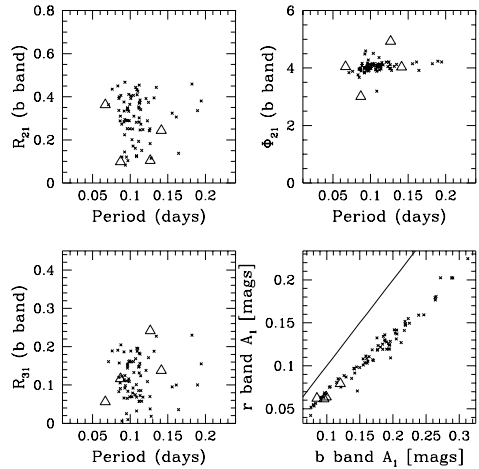}}
\caption{Lightcurve properties based on the Fourier fits to the 4 $\delta$ Scutis detected so
far (open triangles) compared to the \cite{bulge} sample of $\delta$ Scutis in the
Galactic bulge (crosses).  }
\label{four}
\end{figure}

The search thus far has covered all systems from 24 tiles of Field 81
of the MACHO LMC database.  Initial cuts selected systems with
instrumental $r$ magnitudes less than -5 and with colours in the range
of $-0.6 < b-r < 0.2$ and likely variables were selected by their
Welch-Stetson index \citep{wel_stet}.  Periodic variables were
identified by computing the Lomb-normalized periodogram for all
systems using the {\em fasper} algorithm of \cite{press}.  For systems
with periods less than 0.35 days the Discrete Fourier Transform was
computed and the window function removed using the CLEAN algorithm.
Systems possessing peaks with $S/N > 3.0$ were kept.  A Fourier series
was fit to both the $r$ and $b$ light curves of each system and those with
appropriate $R_{21}$ and $\Phi_{21}$ values (31 systems) were
identified as possible $\delta$ Scutis.  Confirmation as $\delta$
Scutis was reserved for those systems whose amplitudes in the $r$ and $b$
bands were in the expected ratio.  Only 4 systems met this final
criterion (open triangles in Fig. \ref{four}).

\section{Discussion}

The four confirmed systems represent only 15\% of the expected number
of HADS based on the local space density calculated from the catalogue
of \citet{rod_bre}.  The use of both bandpasses in selecting $\delta$
Scutis may be too stringent given the amount of scatter in the
lightcurves.

The results presented here represent 24 out of 169 tiles from Field 81
(which contains regions with very young stellar populations).  The
completed survey will also encompass and compare Fields 77 (in the LMC
bar) and 47 (outside the bar).

This work has focussed on the HADS but many other types of variable
star are expected to be recovered.  The most significant at these short
periods a are likely to be eclipsing binaries particularly W UMa type.
\cite{slavek} estimates that W UMa binaries are 8 times less common
than $\delta$ Scutis in the solar neighbourhood but twice as common as
the HADS.

\bibliographystyle{aa}

\begin{thebibliography}{}
\bibitem[Alcock et al.(2000)]{bulge} Alcock, C., et al.\ 
2000, \apj, 536, 798 

\bibitem[Press et al.(1992)]{press} Press, W.~H., Teukolsky, 
S.~A., Vetterling, W.~T., \& Flannery, B.~P.\ 1992, Cambridge: University 
Press, |c1992, 2nd ed.,  
 
\bibitem[Rodr{\'{\i}}guez \& Breger(2001)]{rod_bre} 
Rodr{\'{\i}}guez, E., \& Breger, M.\ 2001, \aap, 366, 178 
 
\bibitem[Rucinski(2004)]{slavek} Rucinski, S.~M.\ 2004, New 
Astronomy Review, 48, 703 
 
\bibitem[Welch \& Stetson(1993)]{wel_stet} Welch, D.~L., \& 
Stetson, P.~B.\ 1993, \aj, 105, 1813 
 
\end{thebibliography}

\end{document}